\documentclass[a4paper,11pt]{article}
\pdfoutput=1

\usepackage{jheppub}

\usepackage{hyperref}
\usepackage{graphicx}
\usepackage{amsmath}
\usepackage{xcolor}
\usepackage{amssymb}
\usepackage[utf8]{inputenc}
\usepackage{subfigure}
\usepackage{bm}
\usepackage{euscript}

\newcommand{\nn}{\nonumber}

\newcommand{\cD}{{\mathcal D}}
\newcommand{\cO}{{\mathcal O}}

\newcommand{\cI}{{\mathcal I}}

\graphicspath{ {figs/} }

\DeclareMathAlphabet{\mathbbold}{U}{bbold}{m}{n}
\allowdisplaybreaks[2]

\preprint{\begin{flushright}
CERN-TH-2024-203
\end{flushright}}

\title{
The two-loop fully differential soft function for $Q\bar{Q}V$ production at lepton colliders
}

\author[a]{Ze Long Liu,}
\emailAdd{liuzelong@ihep.ac.cn}
\author[b]{Pier Francesco Monni}
\emailAdd{pier.monni@cern.ch}

\affiliation[a]{Theoretical Physics Division, Institute of High Energy Physics,
Chinese Academy of Sciences, Beijing 100049, China}
\affiliation[b]{CERN, Theoretical Physics Department, CH-1211 Geneva 23, Switzerland}

\abstract{We consider the production of a pair of heavy quarks $Q\bar{Q}$ in association with a generic colour singlet system $V$ at lepton colliders, and present the first analytic calculation of the two-loop soft function differential in the total momentum of the real radiation. The calculation is performed by reducing the relevant Feynman integrals into a canonical basis of master integrals by means of integration-by-parts identities. The resulting integrals are then evaluated by solving a system of differential equations in the kinematic invariants, whose boundary conditions are determined analytically with some care due to the presence of Coulomb singularities. The fully differential soft function is expressed in terms of Goncharov polylogarithms.
This result is an essential ingredient for a range of N$^3$LL resummations for key collider observables at lepton colliders, such as the $Q\bar{Q}V$ production cross section at threshold and observables sensitive to the total transverse momentum of the radiation in heavy-quark final states. Moreover, it constitutes the complete final-final dipole contribution to the fully differential soft function needed for the description of $Q\bar{Q}V$ production at hadron colliders, which plays an important role in the LHC physics programme.}
\begin{document}
\maketitle

%%%%%%%%%%%%%%%%%%%%%%%%%%%%%%%%%%%%%%%%%%%%%%%%%%%%%%%%%%%%%%%%%%%%%%%%%%%%%%%%
\section{Introduction}

The production of heavy quark pairs ($Q\bar{Q}$), optionally accompanied by a colour singlet system $V$, constitutes a key process at high-energy particle colliders. Both at future lepton colliders and at modern hadron colliders such as the LHC, this class of processes is relevant in the context of the precise extraction of properties of top and bottom quarks and of their couplings with the Higgs and electroweak gauge bosons (see e.g.~\cite{Janot:2015mqv,Janot:2015yza,Vos:2016til,CMS:2021klw,CMS:2022tkv,ATLAS:2023eld,CMS:2024fdo,ATLAS:2024gth,CMS:2024mke,ATLAS:2024moy}).

The theoretical description of physical observables measured on this final state is often plagued by the presence of large logarithmic corrections of infrared origin, that requires a resummation of the perturbative expansion at all orders.
A class of important collider observables admits a factorisation theorem encoding the description of the infrared limit of the observable $v$ at leading power. Key examples of such factorisation theorems at lepton colliders are those describing the production cross section of the $Q\bar{Q}(V)$ system at threshold~\cite{vonManteuffel:2014mva,Gao:2014nva,Gao:2014eea},
\footnote{
The threshold limit discussed in these references corresponds to the energetic heavy quark pair production in association with soft QCD emissions, which is different from the non-relativistic heavy quark pair production, needed to describe accurately the threshold cross section at future lepton colliders~\cite{Hoang:1998xf,Hoang:2013uda,Beneke:2015kwa}.
}
 the back-to-back limit of the energy-energy correlation~\cite{PhysRevD.19.2018,PhysRevLett.41.1585} for heavy-quark final states and, in the case of hadronic collisions, the total transverse momentum of the final-state system with respect to the beam axis~\cite{Zhu:2012ts,Li:2013mia,Catani:2014qha,Catani:2018mei,Ju:2024xhd}.
From a phenomenological viewpoint, such factorisation theorems for this class of processes can be used in the context of slicing methods for higher-order calculations~\cite{Gao:2014nva,Gao:2014eea,Catani:2019iny,Catani:2022mfv,Buonocore:2022pqq} as well as in the matching of next-to-next-to-leading order calculations to parton showers~\cite{Mazzitelli:2020jio,Mazzitelli:2021mmm,Mazzitelli:2024ura}.

An essential ingredient common to all the above factorisation theorems is the soft function, which describes the effect of soft radiation emitted from the eikonalised propagators of the energetic heavy quarks, as encoded in time-like (massive) Wilson lines. 
Interestingly, the soft functions relevant for a broad class of observables, such as the threshold and transverse momentum dependent (TMD) soft function, can be derived starting from a more exclusive soft function differential in the total momentum of the real radiation~\cite{Li:2016axz,Li:2016ctv}. The fully differential soft function with two light-like Wilson lines has been calculated to three loops~\cite{Li:2016ctv}, while the corresponding result with time-like Wilson lines is unknown beyond the one-loop order.
In this article we present the first two-loop calculation of the latter fully differential soft function in the context of the production of a $Q\bar{Q}V$ final state at lepton colliders. This calculation constitutes a key ingredient to improve the description of heavy-quark processes at future lepton colliders, and a consistent building block, that of final-final massive dipoles, for the corresponding calculation for $Q\bar{Q}V$ production at hadron colliders. 
In the hadron-collider case, the two-loop computation of the threshold soft function for $Q\bar{Q}$ production has been carried out in Refs.~\cite{Czakon:2013hxa,Wang:2018vgu}, while numerical calculations have been performed for the corresponding TMD soft function averaged over the azimuthal angle of the final-state system in Refs.~\cite{Angeles-Martinez:2018mqh,Catani:2023tby}. 
The analytic calculation of the fully differential soft function for $Q\bar{Q}V$ production at hadron colliders would be highly desirable. The information encoded in this object would allow for theoretical investigations on the structure of multi-leg soft functions with both light-like and time-like Wilson lines and, from a phenomenological viewpoint, it would be relevant for precision phenomenology involving heavy-quark final states, for instance in the context of higher-order calculations of azimuthal asymmetries~\cite{Catani:2017tuc} which are known to diverge in fixed-order perturbation theory.

The article is structured as follows. In Section~\ref{sec:def} we introduce our notation and conventions, while Section~\ref{sec:calc} outlines the computational method adopted in our calculation up to two loops together with consistency checks. In Section~\ref{sec:renorm} we perform the renormalisation and present our results. Finally, Section~\ref{sec:conclu} contains our conclusions and outlook to future work. The arXiv submission of this article is accompanied by the analytic results for the fully differential soft function distributed as ancillary files.

%%%%%%%%%%%%%%%%%%%%%%%%%%%%%%%%%%%%%%%%%%%%%%%%%%%%%%%%%%%%%%%%%%%%%%%%%%%%%%%%
\section{Definition of the fully differential soft function}
\label{sec:def}
Let us consider the following process in lepton-lepton annihilation
\begin{align}
l^+(\ell_1) + l^-(\ell_2) \to Q_1 (q_1)+ \bar Q_2 (q_2)+ V (q_V) + X (q_X) \,,
\end{align}
where $Q_{1,2}$ are heavy quarks with the same or distinct masses $q_{1,2}^2$, $V$ is a color singlet of momentum $q_V^\mu$, and $X$ denotes the additional QCD radiation of total momentum $q_X^\mu$. Here and below, we use $Q$ and $\varepsilon$ to denote the total energy of the collision and a low scale characterising soft emissions, respectively. At leading power in $\varepsilon/Q$, the cross section can be factorised in the framework of Soft Collinear Effective Theory (SCET)~\cite{Bauer:2000yr,Bauer:2001yt,Bauer:2002nz,Beneke:2002ph,Beneke:2002ni} as follows
\begin{align}\label{eq:xfac}
\frac{d\sigma}{d^4 q \,d\Omega}=\frac{1}{(2\pi)^4}\int d^4 x \ e^{-i (q -q_1 - q_2 - q_V)\cdot x}\ {\cal H}(\{q_1,q_2,q_V\})\ {\cal S}(x,\{v_1, v_2\}) \ + \ {\cal O}(\varepsilon/Q)\,,
\end{align}
where $d\Omega$ indicates the differential phase space element of the $Q_1 Q_2 X$ final state. The hard function ${\cal H}(\{q_1,q_2,q_V\}) $ is the squared of the hard Wilson coefficient (see e.g. Ref.~\cite{Gao:2014nva}), and $\cal S$ is the soft function defined as
\begin{align}\label{eq:sdef}
{\cal S}(x,\{v_1, v_2\})=\frac{1}{N_c} {\rm tr} \left \langle 0 \left | \overline{\bf T}\left[Y_{v_1}(x)Y_{v_2}^\dagger(x)\right]  {\bf T}\left[Y_{v_1}^\dagger(0)Y_{v_2}(0)\right]\right | 0 \right \rangle \,,
\end{align}
where the trace is over color indices, and $Y_{v_i}$ is a soft Wilson line along the velocity $v_i$ of the $i$-th heavy quark 
\begin{align}\label{eq:swilsdef}
Y_{v_i}(x)={\bf P}\exp\left(-ig_s\int_0^\infty dt\ v_i\cdot A_s^a(x +t v_i)\ {\bm T}^a\right)\,.
\end{align}
The cross section given in Eq.~\eqref{eq:xfac} is fully differential on the final-state kinematics, including on the \textit{total} momentum of the real radiation. At leading power, the hard Wilson coefficient is not sensitive to the kinematics of soft emissions in the final state due to the decoupling of hard and soft interactions in the SCET Lagrangian, so $\cal H$ can be taken out from the Fourier transform. Since the cross section in Eq.~\eqref{eq:xfac} is fully exclusive in the radiation's momentum, it can describe observables in various kinematic limits. We present two specific examples below.

The first is the production cross section for $Q_1 Q_2 V$ in the threshold limit, where the total energy of the soft QCD radiations $(q-q_1-q_2-q_V)^0$ is constrained by $E_{\rm cut}\sim \varepsilon\ll Q$. The corresponding threshold soft function can be applied to study the next-to-next-to-leading order (NNLO) QCD corrections to heavy quark pair production in $e^+e^-$ annihilation with a phase-space slicing method~\cite{vonManteuffel:2014mva,Gao:2014nva,Gao:2014eea}. In this case, the spatial coordinate is fixed by $\delta^3(\vec{x})$  after integrating over $d^3\vec{q}$, so in the threshold limit Eq.~(\ref{eq:xfac}) can be rewritten as 
\begin{align}\label{eq:threshold-soft}
\frac{d\sigma(E_{\rm cut})}{d\Omega}={\cal H}(\{q_1,q_2,q_V\}) \int_0^{E_{\rm cut}} d E \int \frac{d x^0}{2\pi}\ e^{-i E\, x^0} {\cal S}\left((x^0,\vec{x}=0),\{v_1,v_2\}\right) \ + \ {\cal O}(E_{\rm cut}/Q)\,.
\end{align}
A second example is that of global observables sensitive to the total transverse momentum of the radiation taken with respect to a reference axis. 
Observables in this class include the energy-energy-correlation~\cite{PhysRevD.19.2018,PhysRevLett.41.1585} for massive quarks, or the transverse momentum of the $Q\bar{Q}V$ system w.r.t. the beam axis. The latter is in direct correspondence with the analogous observable in hadron collisions, of which the calculation presented in this article constitutes a self-consistent building block.
Due to the absence of collinear singularities in the problem under consideration, the specific choice of the reference axis does not lead to additional logarithmic corrections (at leading power in $\varepsilon/Q$). We can therefore parametrise a generic momentum $p^\mu$ using the standard light-cone decomposition
\begin{align}
p^\mu=(n\cdot p)\frac{{\bar n}^\mu}{2} + ({\bar n}\cdot p)\frac{{n}^\mu}{2} + p_{\perp}^\mu \equiv p_{+}^\mu + p_{-}^\mu + p_{\perp}^\mu \,,
\end{align}
where $n$ and $\bar n$ are light-like vectors aligned to the reference directions (e.g. the beam), the total momentum of QCD emissions scales as
\begin{align}
q_X^\mu=(q-q_1-q_2-q_V)^\mu \sim Q (1,1,\varepsilon)\,, \qquad \mbox{with} \quad\varepsilon = q_{X,T}/Q\,,
\end{align}
with $q_{X,T}\equiv \sqrt{-q_{X,\perp}^2}$. 
The conjugate position-space coordinate $x^\mu$ then scales as $x^\mu\sim \left(Q\right)^{-1}(1,1,\varepsilon^{-1})$ where, by performing the multiple-pole expansion for the soft fields, the dependence on light-cone components $x_+$ and $x_-$ are suppressed compared to the perpendicular component $x_\perp$. Therefore, the cross section in Eq.~(\ref{eq:xfac}) can be rewritten as 
\begin{align}\label{eq:xfac-2}
\frac{d\sigma}{d\Omega}= \frac{d^2 q_{X,\perp}}{(2\pi)^2}\ {\cal H}(\{q_1,q_2,q_V\})\int d^2 x_\perp \ e^{-i q_{X,\perp}\cdot x_\perp}\ {\cal S}(x_\perp,\{v_1, v_2\}) \ + \ {\cal O}(q_{X,T}/Q)\,.
\end{align}
As shown in the above examples, the fully differential soft function with the full dependence on $x^\mu$ encodes information that is relevant in a range of factorisation theorems upon appropriate projections.

%%%%%%%%%%%%%%%%%%%%%%%%%%%%%%%%%%%%%%%%%%%%%%%%%%%%%%%%%%%%%%%%%%%%%%%%%%%%%%%%
\section{Calculation of the two-loop fully differential soft function}\label{sec:calc}
In the present section we will outline the main aspects of the calculation. We first introduce the setup and computational strategy, and then discuss both the NLO and NNLO results.

\subsection{Fully differential soft function in momentum space}\label{sec:smom}
Eq.~(\ref{eq:sdef}) defines the soft function in position space. Due to the convenience of evaluation of Feynman integrals in momentum space, it makes sense to perform the calculation in momentum space first, and then Fourier-transform the result back to position space. The fully differential soft function in momentum space can be defined as
\begin{align}\label{eq:sdef-momentum-space}
S(\omega,\{\eta,v_1, v_2\})=\frac{1}{N_c} {\rm tr} \left \langle 0 \left | \overline{\bf T}\left[Y_{v_1}(0)Y_{v_2}^\dagger(0)\right]\delta(\omega-\eta\cdot \hat{p})\ {\bf T}\left[Y_{v_1}^\dagger(0)Y_{v_2}(0)\right]\right | 0 \right \rangle \,,
\end{align}
where $\eta^\mu$ is a dimensionless vector aligned with $x^\mu$, and ${\hat {p}}^\mu$ is an operator picking up the total momentum of soft radiations in the final state. With a Fourier transform, $S(\omega,\{\eta,v_1, v_2\})$ can be related to the soft function in position space
\begin{align}\label{eq:sxtop}
{\cal S}(x,\{v_1,v_2\})=\int d\omega\ e^{i\omega t}\ S(\omega,\{\eta,v_1, v_2\})\,,
\end{align}
where $t$ relates $\eta^\mu$ to $x^\mu$ by $x^\mu = t\eta^\mu$. The explicit $\eta^\mu$ and $t$ for threshold and TMD soft functions can be chosen as
\begin{align}
&\eta^\mu = (1,0,0,0)\,,  \qquad\quad t=x^0\,, \qquad \mbox{and}\label{eq:threshold-case} \\
&\eta^\mu = (0,-i \cos \phi,-i \sin \phi,0)\,,  \qquad t=i\, x_T\,\label{eq:tmd-case}, 
\end{align}    
respectively, where $\phi$ is the azimuthal angle between $p_\perp$ and $x_\perp$, and $x_T\equiv \sqrt{-x_\perp^2}$. 
%
%In practice, it is more convenient to evaluate the TMD soft function in momentum space in the rest frame of $\eta^\mu$. 
%
We stress that the fully differential soft function considered here is free of rapidity divergences. In the more general case of a fully differential soft function containing also light-like Wilson lines, one should supplement the correspondence given in Eq.~\eqref{eq:tmd-case} with a rapidity regularisation procedure. Within the exponential regularisation scheme of Ref.~\cite{Li:2016axz}, this would simply amount to adding a real time component to the vector $\eta^\mu$ in Eq.~\eqref{eq:tmd-case}.

Since the soft Wilson lines in Eq.~(\ref{eq:sdef-momentum-space}) are invariant under rescaling $v_1^\mu\to \alpha_1 v_1^\mu$ and $v_2^\mu\to \alpha_2 v_2^\mu$, the soft function depends on $v_1^\mu$ and $v_2^\mu$ only through cross ratios like $(v_1\cdot v_2)/(\sqrt{v_1^2}\sqrt{v_2^2})$. In our case, there are only three independent cross ratios involved in the soft function, which can be chosen as (with a little abuse of notation we denote by $x$ one of these cross ratios, this is not to be confused with the coordinate $x^\mu$ used as argument of the fully differential soft function)
\begin{gather}\label{eq:xratios}
x=\frac{v_1\cdot v_2}{\sqrt{v_1^2}\sqrt{v_2^2}}\,, \qquad
y=\frac{v_1\cdot \eta}{\sqrt{v_1^2}\sqrt{\eta^2}}\,, \qquad
z=\frac{v_2\cdot \eta}{\sqrt{v_2^2}\sqrt{\eta^2}}\,.
\end{gather}
Therefore, by dimensional counting in dimensional regularisation, the bare soft function  takes the following generic form at $n$ loops
\begin{align}
S^{(n)}(\omega,\{\eta,v_1, v_2\};\epsilon) 
= \omega^{-1-2n\epsilon}\left(\frac{\eta^2}{4}\right)^{n\epsilon} f_n(x,y,z;\epsilon) \,,
\end{align}
where $\epsilon=(4-d)/2$ denotes the dimensional regulator, and we use the notation $S(\omega,\{\eta,v_1, v_2\};\epsilon) =\delta(\omega) + \sum_{n=1} \left(\frac{\alpha_s}{4\pi}\right)^n S^{(n)}(\omega,\{\eta,v_1, v_2\};\epsilon)$. 

%%%%%%%%%%%%%%%%%%%%%%%%%%%%%%%%%%%%%%%%%%%%
\subsection{NLO calculation}\label{sec:nlo}
Since purely virtual corrections are expressed by scaleless integrals, the NLO correction to the soft function only receives contributions from single-real radiation, which can be expressed as
\begin{align}
S^{(1)}(\omega;\epsilon) 
= C_\epsilon^{(1)} C_F\int d^d k\ \delta^+(k^2)\ \delta(\omega - \eta\cdot k)
\left[\frac{v_1^2}{(v_1\cdot k)^2}
+\frac{v_2^2}{(v_2\cdot k)^2}
-\frac{2v_1\cdot v_2}{(v_1\cdot k)(v_2\cdot k)}
\right]\,,
\end{align}
where $C_\epsilon^{(1)}=2 e^{\epsilon\gamma_E} \pi^{\epsilon -1}$. Here and below, we omit the arguments $\{\eta,v_1, v_2\}$ in the soft function for simplicity. Because of the complexity of the kinematics, it is still challenging to evaluate the one-loop phase space integrals directly. To accomplish analytical computation, we employ integration by parts (IBP) reduction~\cite{Tkachov:1981wb,Chetyrkin:1981qh} and the method of differential equation (DE)~\cite{Kotikov:1990kg,Kotikov:1991hm,Kotikov:1991pm,Bern:1992em,Bern:1993kr,Remiddi:1997ny,Gehrmann:1999as}. To perform IBP reduction for phase space integrals, we need to replace the $\delta$ functions with propagators by means of reverse unitary~\cite{Anastasiou:2002yz,Anastasiou:2003yy,Anastasiou:2003ds}
\begin{align}
\delta(x) = \frac{1}{2\pi i}\left(\frac{1}{x-i0}-\frac{1}{x+i0}\right)
\equiv \mbox{Disc}(x^{-1})\,.
\end{align}
Then we rewrite $S^{(1)}(\omega;\epsilon)$ as
\begin{align}\label{eq:amp1l}
S^{(1)}(\omega;\epsilon) 
= C_F \ \omega^{-1-2\epsilon} \left(\frac{\eta^2}{4}\right)^\epsilon \left[v_1^2\ I_{1,1,2,0}+ v_2^2\ I_{1,1,0,2}- 2(v_1\cdot v_2)\ I_{1,1,1,1}\right]\,,
\end{align}
with
\begin{align}\label{eq:integrals-nlo}
I_{a_1,a_2,a_3,a_4}
= \kappa_1(\{a_i\})\ C_\epsilon^{(1)} \int \! d^d k\ \mbox{Disc}\left[(k^2)^{-a_1}\right]
\mbox{Disc}\left[(\omega-\eta\cdot k)^{-a_2}\right]
\frac{(-1)^{a_3+a_4}}{(v_1\cdot k)^{a_3}(v_2\cdot k)^{a_4}}\,,
\end{align}
where the normalisation factor 
\begin{align}
 \kappa_1(\{a_i\})=\omega^{2a_1+a_2+a_3+a_4-d}
 \left(\frac{\eta^2}{4}\right)^{(d-2a_1-a_3-a_4)/2} \nn
\end{align}
eliminates the dependence on $\omega$ and $\eta^2$.\footnote{$I_{a_1,a_2,a_3,a_4}$ depends on $\eta^\mu$ only through the cross ratios in Eq.~(\ref{eq:xratios}).} 
The parametrisation of the phase-space integrals given in Eq.~\eqref{eq:integrals-nlo} allows one to use standard Feynman integrals technology for their evaluation. 
By using the package \texttt{FIRE6}~\cite{Smirnov:2019qkx} for IBP reduction, $S^{(1)}(\omega;\epsilon)$ is expressed in terms of a set of master integrals (MIs). Their derivatives with respect to the cross ratios can be obtained with the package \texttt{LiteRed}~\cite{Lee:2013mka}.
To solve for the MIs, we choose the following basis of MIs with uniform transcendental weight (UT)    
\begin{align}\label{eq:1lMIs}
 {\vec \cI}=\{\,
 2(1-2 \epsilon)\, I_{1,1,0,0}\,,\
 \epsilon\sqrt{1-1/z^2} \, I_{1,1,0,1}\,,\ 
 \epsilon\sqrt{1-1/y^2} \, I_{1,1,1,0}\,,\ 
 \epsilon\sqrt{1-1/x^2} \, I_{1,1,1,1}\}\,,   
\end{align}
which leads to a simple DE system with $\epsilon$-form~\cite{Henn:2013pwa}
\begin{align}\label{eq:de1l}
\partial_x\vec{\cI}=&\epsilon\left(
\begin{array}{cccc}
 0 & 0 & 0 & 0 \\
 0 & 0 & 0 & 0 \\
 0 & 0 & 0 & 0 \\
 \frac{2}{\sqrt{x^2-1}} & \frac{-4 \sqrt{z^2-1} (z-x
   y)}{\sqrt{x^2-1} \left(x^2-2 x y z+y^2+z^2-1\right)}
   & \frac{-4 \sqrt{y^2-1} (y-x z)}{\sqrt{x^2-1}
   \left(x^2-2 x y z+y^2+z^2-1\right)} & \frac{2
   \left(\left(x^2+1\right) y z-x y^2-x
   z^2\right)}{\left(x^2-1\right) \left(x^2-2 x y
   z+y^2+z^2-1\right)} \\
\end{array}
\right) \cdot \vec{\cI}\,,  \nn 
\\
\nn\\
\partial_y\vec{\cI}=&\epsilon\left(
\begin{array}{cccc}
 0 & 0 & 0 & 0 \\
 0 & 0 & 0 & 0 \\
 -\frac{1}{\sqrt{y^2-1}} & 0 & \frac{2 y}{y^2-1} & 0 \\
 0 & -\frac{4 \sqrt{x^2-1} \sqrt{z^2-1}}{x^2-2 x y
   z+y^2+z^2-1} & \frac{4 \sqrt{x^2-1} (y
   z-x)}{\sqrt{y^2-1} \left(x^2-2 x y
   z+y^2+z^2-1\right)} & \frac{2 (y-x z)}{x^2-2 x y
   z+y^2+z^2-1}\\
\end{array}
\right)\cdot \vec{\cI} \,, 
\\
\nn\\
\partial_z\vec{\cI}=&\epsilon\left(
\begin{array}{cccc}
 0 & 0 & 0 & 0 \\
 -\frac{1}{\sqrt{z^2-1}} & \frac{2 z}{z^2-1} & 0 & 0 \\
 0 & 0 & 0 & 0 \\
 0 & \frac{4 \sqrt{x^2-1} (y z-x)}{\sqrt{z^2-1}
   \left(x^2-2 x y z+y^2+z^2-1\right)} & -\frac{4
   \sqrt{x^2-1} \sqrt{y^2-1}}{x^2-2 x y z+y^2+z^2-1} &
   \frac{2 (z-x y)}{x^2-2 x y z+y^2+z^2-1} \\
\end{array}
\right)\cdot \vec{\cI} \,.\nn
\end{align}
The boundary conditions for solving the DEs are determined by the values of MIs at the phase space point $\eta^\mu=v_1^\mu=v_2^\mu$. These are obtained by recasting the denominators that were processed with reverse unitarity (see e.g. Eq.~\eqref{eq:integrals-nlo}) as delta functions, and performing the corresponding phase space integrals. They are given by
\begin{align}
\vec{\cI}_0=\left\{
\frac{2 e^{\epsilon\gamma_E} \Gamma (1-\epsilon)}{\Gamma (1-2 \epsilon )},0,0,0
\right\}\,.
\end{align}
By rationalizing the square roots in Eq.~(\ref{eq:de1l}) with the change of variables
\begin{align}\label{eq:ratvars}
 r = \sqrt{x^2-1}-x+1 \,,\qquad
 t = \sqrt{y^2-1}-y+1 \,,\qquad
 u = \sqrt{z^2-1}-z+1 \,,
\end{align}
and solving the DEs order by order in $\epsilon$, the MIs are expressed in terms of Goncharov polylogarithms (GPLs), which can be defined iteratively as
\begin{align}
    G(a_1,\dots\,a_n;x)=\int_0^x\frac{dt}{t-a_1}G(a_2,\dots\,a_n;x)\,,
\end{align}
with $G(;x)=1$ and $G(\vec{0}_n;x)=1/n!\,\ln^n x$\,. At NLO, the letters involved in the arguments of GPLs are $\{0,1,2,\lambda_1,\lambda_2,\lambda_3,\lambda_4\}$, where $\lambda_{i=1,\dots ,4}$ are the roots of the polynomial $ \lambda^4 + \lambda^3 (4 y z-4)  + 4 \lambda^2 \left(y^2-3 y z+z^2+1\right)-8\lambda (y-z)^2+4 (y-z)^2 =0 \,$, and take the values
\begin{align}\label{eq:letts1l}
\lambda_1 = & 1 -\sqrt{\left(y^2-1\right) \left(z^2-1\right)}-\sqrt{2 y^2 z^2 +2 y z \sqrt{\left(y^2-1\right) \left(z^2-1\right)} -y^2 - z^2}-y z \,,\nn \\
\lambda_2 = & 1 -\sqrt{\left(y^2-1\right) \left(z^2-1\right)}+\sqrt{2 y^2 z^2 +2 y z \sqrt{\left(y^2-1\right) \left(z^2-1\right)} -y^2 - z^2}-y z \,,\nn \\
\lambda_3 = & 1 - \sqrt{\left(y^2-1\right) \left(z^2-1\right)}-\sqrt{2 y^2 z^2 -2 y z
   \sqrt{\left(y^2-1\right) \left(z^2-1\right)} -y^2 - z^2}-y z \,,\nn\\
\lambda_4 = & 1 - \sqrt{\left(y^2-1\right) \left(z^2-1\right)}+\sqrt{2 y^2 z^2 -2 y z
   \sqrt{\left(y^2-1\right) \left(z^2-1\right)} -y^2 - z^2}-y z  \,. 
\end{align}
The result of NLO bare soft function will be presented in Sec.~\ref{sec:renorm}.
%%%%%%%%%%%%%%%%%%%%
\subsection{NNLO calculation}
%%%%%%%%%%%%%%%%%%%%%%%%%%%%%%%%%%%%%%%%%%
There are two types of contributions to the soft function at NNLO. One is from double real emissions (RR), and the other is from a single real emission with a one-loop virtual correction (RV). In principle, the RV contribution can be obtained by evaluating one-loop phase space integrals for the one-loop massive soft current~\cite{Bierenbaum:2011gg,Czakon:2018iev}. However, it is quite challenging to perform the phase space integration without IBP reduction and the method of DE due to the complexity arising from the measurement of the fully differential soft function. Hence, we follow the same strategy to calculate the RR and RV NNLO contributions as that used for the NLO calculation.  

\begin{figure}[t]
	\begin{center}
		\subfigure[]{\label{fig:triplea}\includegraphics[width=0.21\textwidth]{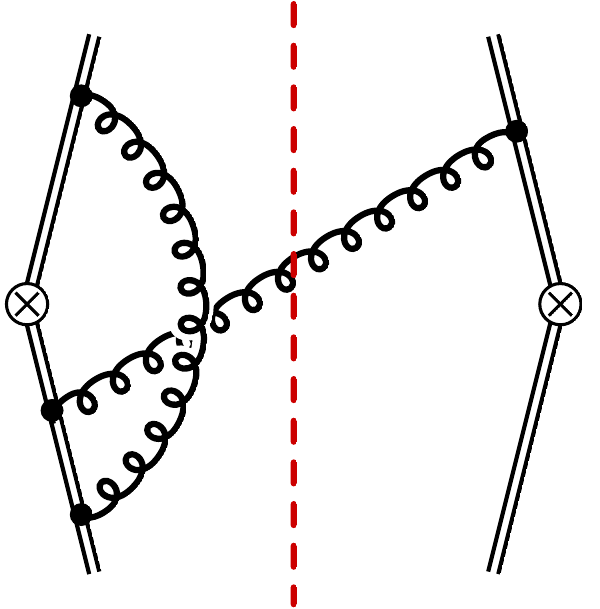}}\hspace{1.6cm}
		\subfigure[]{\label{fig:tripleb}\includegraphics[width=0.21\textwidth]{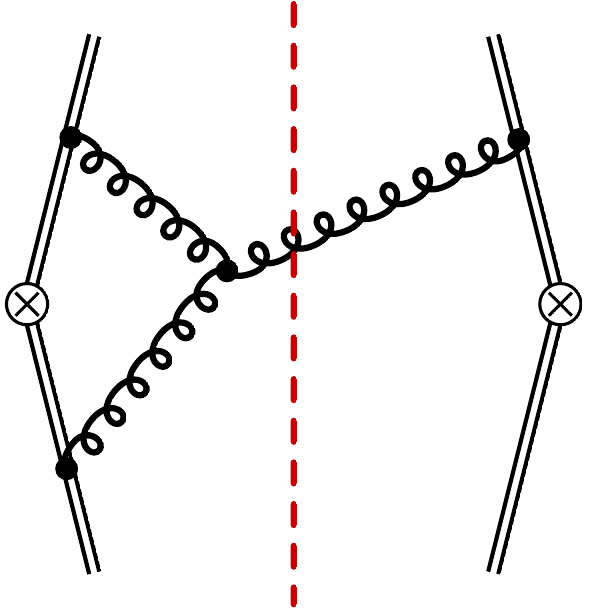}}\hspace{1.6cm}
		\subfigure[]{\label{fig:triplec}\includegraphics[width=0.21\textwidth]{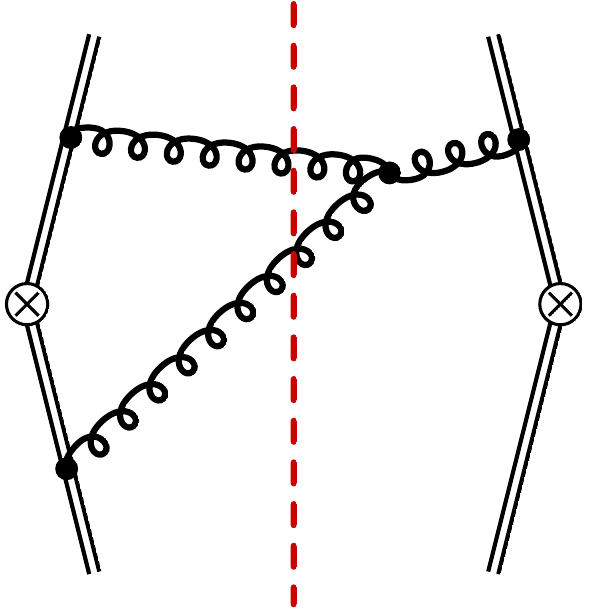}}\\
		\subfigure[]{\label{fig:triplec}\includegraphics[width=0.21\textwidth]{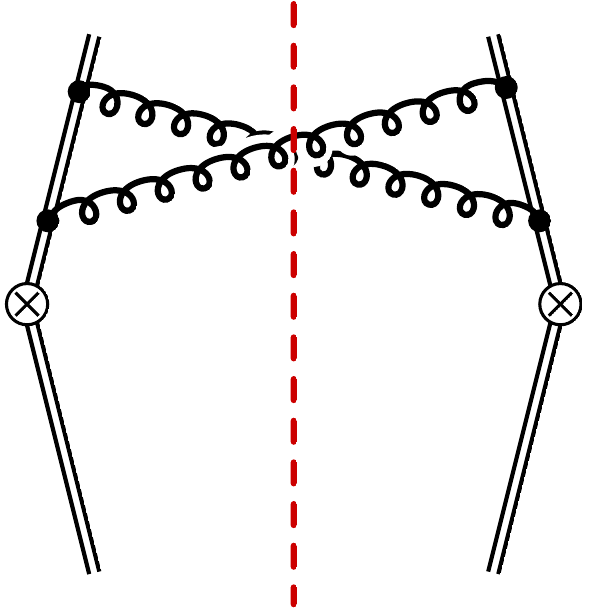}}\hspace{1.6cm}
        \subfigure[]{\label{fig:triplec}\includegraphics[width=0.21\textwidth]{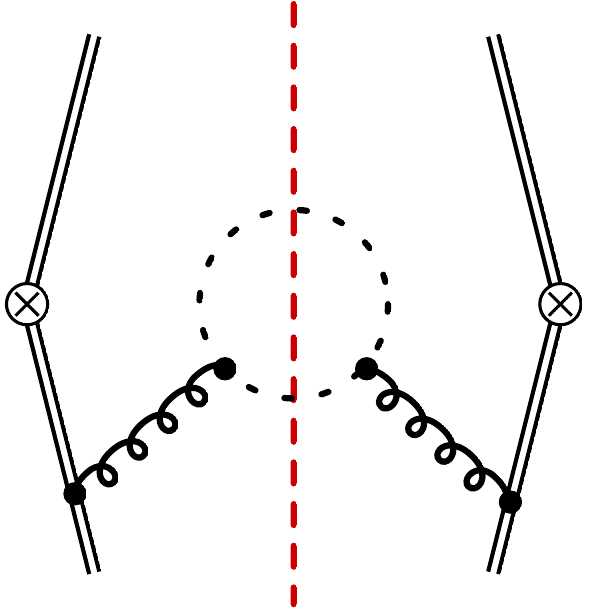}}\hspace{1.6cm}
        \subfigure[]{\label{fig:triplec}\includegraphics[width=0.21\textwidth]{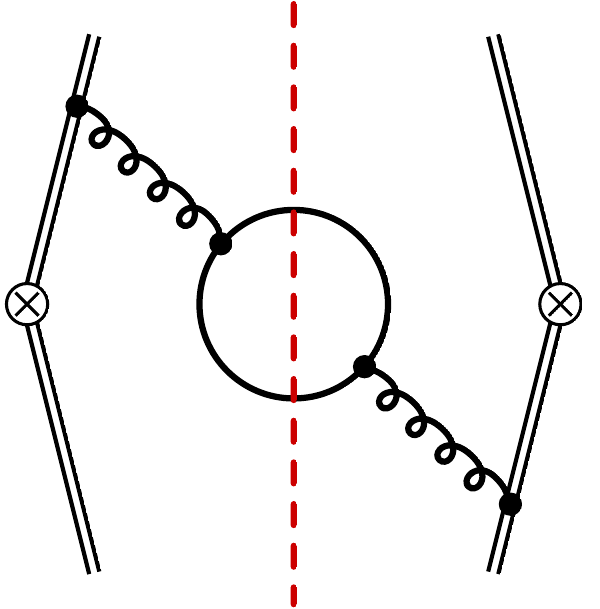}}
	\end{center}
	\caption{\label{fig:feyndia} Examples for two-loop Feynman diagrams that contribute to the soft function $S(\omega;\epsilon)$.}
\end{figure}

To obtain the amplitudes, we generate the two-loop Feynman diagrams with \texttt{qgraf} ~\cite{Nogueira:1991ex} for both the RR and RV contributions individually.
%.
Some examples of two-loop graphs are shown in fig.~\ref{fig:feyndia}. The Feynman gauge is employed for summing over gluon polarisation vectors, so the contribution of ghost fields to the vacuum polarisation diagrams has to be taken into account to cancel the contribution of unphysical gluon polarisations. After assigning the Feynman rules in momentum space and performing Dirac, Lorentz and color algebras, the amplitudes of the RR and RV contributions are expressed as linear combinations of scalar Feynman integrals with Cutkosky cuts. By means of partial-fraction decomposition for linear propagators followed by suitable shifts of the loop momentum, the scalar integrals can be mapped onto the integral topologies defined by complete bases of propagators, which fit to the setup of $\texttt{FIRE6}$.

For the RR contribution, the scalar integrals take the generic form
\begin{align}\label{eq:IRR}
I^{{\rm RR}}_{\vec{a}}
=\kappa_2(\{a_i\})\ C_\epsilon^{\rm RR} \int \! d^d k_1 \int \! d^d k_2\, 
&\mbox{Disc}\left[\left(k_1^2\right)^{-a_1}\right]
\mbox{Disc}\left[\left(k_2^2\right)^{-a_2}\right]\mbox{Disc}\left[(\omega-\eta\cdot (k_1+k_2))^{-a_3}\right]
\nn\\
\times& 
\frac{(-1)^{a_4+a_5+a_6+a_7+a_8}}{\left[(k_1+k_2)^2\right]^{a_4}\cD_5^{a_5}\,\cD_6^{a_6}\,\cD_7^{a_7}\,\cD_8^{a_8}}\,,
\end{align}
where the normalisation factors read
\begin{align}
 \kappa_2(\{a_i\})=\omega^{2a_1+2a_2+a_3+2a_4+a_5+a_6+a_7+a_8-2d}
 \left(\frac{\eta^2}{4}\right)^{(2d-2a_1-2a_2-a_3-2a_4-a_5-a_6-a_7-a_8)/2}\,,
\end{align}
and $C_\epsilon^{\rm RR}=-4 \pi ^{2 \epsilon -2}$. 
As for the NLO case, the reverse unitarity representation of the phase-space integrals is used for the IBP reduction and to derive the canonical differential equations. The evaluation of the boundary conditions is then carried out by re-expressing the relevant denominators as delta functions.
The denominators $\cD_{5,6}$ and $\cD_{7,8}$ correspond to propagators of Wilson lines along the $v_1$ and $v_2$ directions, respectively. After performing IBP reduction and remapping the MIs across the initial integral topologies with \texttt{Kira}~\cite{Klappert:2020nbg}, the RR amplitudes can be expressed by a linear combination of 31 MIs, which can be cast into the following five integral topologies defined by propagators $\{\cD_5,\cD_6,\cD_7,\cD_8\}$ as
\begin{align}
\mbox{Topo RR1 :} &\qquad  \left\{ v_1\cdot k_1\,, v_1\cdot k_2\,, v_2\cdot k_1\,, v_2\cdot k_2 \right\}  \,,
\nn\\
\mbox{Topo RR2 :} &\qquad  \left\{ v_1\cdot k_1 \,, v_1\cdot \left(k_1+k_2\right) \,, v_2\cdot k_1 \,, v_2\cdot  k_2 \right\}  \,,
\nn\\
\mbox{Topo RR3 :} &\qquad  \left\{ v_1\cdot k_1\,, v_1\cdot k_2 \,, v_2\cdot k_1\,, v_2\cdot \left(k_1+k_2\right) \right\}  \,,
\\
\mbox{Topo RR4 :} &\qquad  \left\{ v_1\cdot k_1\,, v_1\cdot \left(k_1+k_2\right)\,, v_2\cdot k_1\,, v_2\cdot \left(k_1+k_2\right)  \right\}  \,,
\nn\\
\mbox{Topo RR5 :} &\qquad  \left\{ v_1\cdot k_1\,, v_1\cdot \left(k_1+k_2\right)\,, v_2\cdot k_2\,, v_2\cdot \left(k_1+k_2\right)  \right\}  \,.\nn
\end{align}
For double-real integrals, the causal `$i0$' prescriptions for the propagators can be neglected, because the denominators are always positive for on-shell momenta $k_{1,2}$. To build the DE system in $\epsilon$-form, we first find the candidate UT integrals by requiring that they have constant leading singularities~\cite{Cachazo:2008vp,Arkani-Hamed:2010pyv}. The method of Magnus and Dyson series expansion~\cite{Argeri:2014qva} is also employed to convert the inhomogeneous terms of the DE into $\epsilon$-form. Eventually, the UT basis of the MIs reads
\begin{align}
\vec{\cI}&^{\rm RR}= \Big\{
\epsilon I_{2,2,1,0,0,0,0,0}^{{\rm RR}1}\,, ~
\frac{\epsilon^2 \sqrt{z^2-1}}{z} I_{1,2,1,0,0,0,1,0}^{{\rm RR}1}\,, ~
\frac{\epsilon^2 \sqrt{y^2-1}}{y} I_{1,2,1,0,1,0,0,0}^{{\rm RR}1}\,, ~
\frac{\epsilon^2 \sqrt{y^2-1}}{y} I_{1,2,1,0,0,1,0,0}^{{\rm RR}2}\,,
\nn\\&
\frac{\epsilon^2 \sqrt{z^2-1}}{z} I_{1,2,1,0,0,0,0,1}^{{\rm RR}3}\,, ~
\frac{(1-4 \epsilon) \epsilon^2 \left(z^2-1\right)}{z^2} I_{1,1,1,0,0,0,1,1}^{{\rm RR}1}\,, ~
\frac{(1-4 \epsilon) \epsilon^2 \sqrt{y^2-1} \sqrt{z^2-1}}{y z} I_{1,1,1,0,0,1,1,0}^{{\rm RR}1}\,,
\nn\\&
\frac{\epsilon^2 \sqrt{x^2-1}}{x}  I_{1,2,1,0,1,0,1,0}^{{\rm RR}1}\,, ~
\frac{(1-4 \epsilon) \epsilon^2 \left(y^2-1\right)}{y^2} I_{1,1,1,0,1,1,0,0}^{{\rm RR}1}\,, ~
\frac{\epsilon^2 \sqrt{y^2-1}}{x z} I_{1,1,1,0,0,1,2,0}^{{\rm RR}2}\,,
\nn\\&
\frac{\epsilon^2 (4 \epsilon-1) \left(x^2-2 y z x+z^2\right)}{x y z} I_{1,1,1,0,0,1,1,0}^{{\rm RR}2}
-\frac{\epsilon^2 (x y-z)}{2 x^2 z} I_{1,1,1,0,0,1,2,0}^{{\rm RR}2}
+ \frac{\epsilon^2(x y-z)}{x y} I_{1,2,1,0,0,1,0,0}^{{\rm RR}2}\,,
\nn\\&
\frac{(1-4 \epsilon) \epsilon^2 \left(y^2-1\right)}{y^2} I_{1,1,1,0,1,1,0,0}^{{\rm RR}2}\,, ~
\frac{\epsilon x}{\sqrt{x^2-1}}  I_{2,2,1,0,0,0,0,0}^{{\rm RR}1}
+\frac{8\epsilon^2 (4\epsilon-1)\left(x^2-2 y z x+z^2\right)}{y z\sqrt{x^2-1} } I_{1,1,1,0,0,1,1,0}^{{\rm RR}2}
\nn\\&
\qquad +\frac{8 \epsilon^2 (z-x y)}{x z\sqrt{x^2-1} }  I_{1,1,1,0,0,1,2,0}^{{\rm RR}2}
+\frac{8\epsilon^2 (x y-z)}{y \sqrt{x^2-1} } I_{1,2,1,0,0,1,0,0}^{{\rm RR}2}
+\frac{\epsilon (2 \epsilon+1)}{x \sqrt{x^2-1}} I_{1,2,1,0,0,1,0,1}^{{\rm RR}2}\,,
\nn\\&
\frac{(1-4 \epsilon) \epsilon^2 \left(z^2-1\right)}{z^2} I_{1,1,1,0,0,0,1,1}^{{\rm RR}3}\,, ~
\frac{\epsilon^2 \sqrt{z^2-1}}{x y} I_{1,1,1,0,2,0,0,1}^{{\rm RR}3}\,,
\nn\\&
\frac{\epsilon^2 (4 \epsilon-1) \left(x^2-2 y z x+y^2\right)}{x y z} I_{1,1,1,0,1,0,0,1}^{{\rm RR}3}
 +\frac{\epsilon^2 (y-x z)}{2 x^2 y} I_{1,1,1,0,2,0,0,1}^{{\rm RR}3}
 +\frac{\epsilon^2 (x z - y)}{x z} I_{1,2,1,0,0,0,0,1}^{{\rm RR}3}\,,
\nn\\&
\frac{\epsilon x}{\sqrt{x^2-1}} I_{2,2,1,0,0,0,0,0}^{{\rm RR}1}
+\frac{8\epsilon^2 (4\epsilon-1)\left(x^2-2 y z x+y^2\right)}{ y z \sqrt{x^2-1}} I_{1,1,1,0,1,0,0,1}^{{\rm RR}3}
+\frac{8 \epsilon^2 (y-x z)}{x y\sqrt{x^2-1} } I_{1,1,1,0,2,0,0,1}^{{\rm RR}3} 
\nn\\&\qquad
+\frac{8 \epsilon^2 (x z-y)}{ z \sqrt{x^2-1}} I_{1,2,1,0,0,0,0,1}^{{\rm RR}3}
+\frac{\epsilon (2 \epsilon+1)}{x \sqrt{x^2-1}} I_{1,2,1,0,0,1,0,1}^{{\rm RR}3}\,, ~
\frac{\epsilon^2 \sqrt{x^2-1}}{x} I_{1,2,1,0,0,1,0,1}^{{\rm RR}4}\,,
\nn\\&
\frac{\epsilon^3 \sqrt{x^2-1} \sqrt{z^2-1}}{x z} I_{1,1,1,0,1,0,1,1}^{{\rm RR}1}\,, ~
\frac{\epsilon^3 \sqrt{x^2-1} \sqrt{y^2-1}}{x y} I_{1,1,1,0,1,1,1,0}^{{\rm RR}1}\,, ~
\frac{\epsilon^3 \sqrt{x^2-1} \sqrt{y^2-1}}{x y} I_{1,1,1,0,1,1,0,1}^{{\rm RR}2}\,,
\nn\\&
\frac{\epsilon^3 \sqrt{x^2-1} \sqrt{y^2-1}}{x y} I_{1,1,1,0,1,1,1,0}^{{\rm RR}2}\,, ~
\frac{\epsilon^3 \sqrt{x^2-1}}{x} I_{1,1,1,1,1,0,0,1}^{{\rm RR}2}\,, ~
\frac{\epsilon^3 \sqrt{y^2-1}}{y} I_{1,1,1,1,1,0,-1,1}^{{\rm RR}2}\,,
\nn\\&
\frac{\epsilon^3 \sqrt{z^2-1}}{z} I_{1,1,1,1,1,-1,0,1}^{{\rm RR}2}\,, ~
4 \epsilon^2 I_{1,2,1,0,1,0,0,0}^{{\rm RR}1}
+8 \epsilon^3 I_{1,1,1,1,1,0,-1,1}^{{\rm RR}2}
+2 \epsilon^3 I_{1,1,1,1,1,0,0,1}^{{\rm RR}2}
+\frac{\epsilon^2 z}{x y} I_{1,1,1,1,2,0,-1,1}^{{\rm RR}2}\,,
\nn\\&
\frac{\epsilon^3 \sqrt{x^2-1} \sqrt{z^2-1}}{x z} I_{1,1,1,0,0,1,1,1}^{{\rm RR}3}\,, ~
\frac{\epsilon^3 \sqrt{x^2-1} \sqrt{z^2-1}}{x z} I_{1,1,1,0,1,0,1,1}^{{\rm RR}3}\,, ~
\frac{\epsilon^3 \left(x^2-1\right)}{x^2} I_{1,1,1,0,1,1,1,1}^{{\rm RR}1}\,,
\nn\\&
\frac{\epsilon^3 \left(x^2-1\right)}{x^2} I_{1,1,1,0,1,1,1,1}^{{\rm RR}4}\,, ~
\frac{\epsilon^3 \left(x^2-1\right)}{x^2} I_{1,1,1,0,1,1,1,1}^{{\rm RR}5}
\Big\}\,,
\end{align}
where the normalisation factors are obtained from the evaluation of the leading singularities of corresponding integrals in Baikov representation~\cite{Baikov:1996rk,Baikov:2005nv}. 
The boundary conditions are determined at the point $v_1^\mu=v_2^\mu=\eta^\mu$. By using the Mellin-Barnes technique~\cite{Smirnov:1999gc,Tausk:1999vh} implemented in the $\texttt{MB}$ package~\cite{Czakon:2005rk,Smirnov:2009up}, we obtain the analytical results for the boundary conditions as 
\begin{align}
   \cI^{\rm RR}_{1,26}\Big |_{x=y=z=1} & = 8-\frac{28 \pi ^2 \epsilon ^2}{3}-\frac{496 \zeta_3 \epsilon ^3}{3}-\frac{\pi ^4 \epsilon ^4}{5}+\cO(\epsilon^5)\,, \\
   \cI^{\rm RR}_{i}|_{x=y=z=1} &= 0 \qquad \mbox{for}\quad i\neq 1~\mbox{and}~26 \,. \nn
\end{align}
After performing the rationalisation of the square roots shown in Eq.~(\ref{eq:ratvars}) and solving the DE equations order by order in $\epsilon$, the MIs can be expressed in terms of GPLs. The corresponding letters involved in the arguments of the GPLs are 
$\{0\,, 1\,, 2\,, 1-y\,, (y-1)/y\,, \lambda_{i=1,\dots ,18} \}$. $\lambda_{i=1,\dots ,4}$ are the same as the letters in Eq.~(\ref{eq:letts1l}), while $\lambda_{i=5,\dots ,8}$, $\lambda_{i=9,\dots ,12}$, $\lambda_{i=13,\dots ,16}$ and $\lambda_{i=17,18}$ are the roots of the polynomial equations
\begin{align}
& \lambda ^4 + \lambda ^3 (4 y z-4)  + 4 \lambda ^2 \left(z^2-3 y z+2\right) -8\lambda\left(z^2 -2 y z+1 \right) + 4 \left(z^2-2 y z+1\right)   =0 \,,
\nn\\
& \lambda ^4 + \lambda ^3 (4 y z-4)  + 4 \lambda ^2 \left(y^2-3 y z+2\right) -8\lambda\left(y^2-2 y z+1\right) + 4 \left(y^2-2 y z+1\right)  = 0 \,,
\nn\\
& \lambda ^4+4 \lambda ^3 (y-1)-12 \lambda ^2 (y-1)+16 \lambda  (y-1)-8 (y-1)  = 0 \,,
\nn\\
& \lambda ^2+2 \lambda  (y-1)-2 (y-1) = 0 
\,,
\end{align}
respectively.

The RV contribution is addressed with a similar procedure. The real-virtual integrals can be written as
\begin{align}\label{eq:IRV}
I^{{\rm RV}}_{\vec{a}}
=\kappa_2(\{a_i\})\ C_\epsilon^{\rm RV} \int& \! d^d k_1 \, 
\mbox{Disc}\left[\left(k_1^2\right)^{-a_1}\right] 
\mbox{Disc}\left[(\omega-\eta\cdot k_1)^{-a_3}\right]
\nn\\
\times& \int \! d^d k_2\,
\frac{(-1)^{a_2+a_4+a_5+a_6+a_7+a_8}}{\left(k_2^2+i0\right)^{a_2} \left[(k_1+k_2)^2+i0\right]^{a_4}\cD_5^{a_5}\,\cD_6^{a_6}\,\cD_7^{a_7}\,\cD_8^{a_8}}\,,
\end{align}
with $C_\epsilon^{\rm RV}  = 2 i \pi ^{2 \epsilon -3} e^{2\gamma_E}$. After IBP reduction, the amplitudes can be expressed by a linear combination of 20 MIs, which can be remapped into three integral topologies defined by the propagators $\{\cD_5\,,\cD_6\,,\cD_7\,,\cD_8\}$ as 
\begin{align}
\mbox{Topo RV1 :} &\qquad  \left\{ v_1\cdot k_1 \,, - v_1\cdot k_2+i0 \,, v_2\cdot k_1 \,, - v_2\cdot k_2+i0 \right\} \,,
\nn\\
\mbox{Topo RV2 :} &\qquad  \left\{ v_1\cdot k_1 \,, - v_1\cdot k_2+i0 \,, v_2\cdot k_2+i0 \,,  v_2\cdot \left(k_1+k_2\right)+i0 \right\} \,,
\nn\\
\mbox{Topo RV3 :} &\qquad  \left\{ v_1\cdot k_2+i0 \,, v_1\cdot \left(k_1+k_2\right)+i0 \,, v_2\cdot k_1 \,, - v_2\cdot k_2+i0 \right\} \,.
\end{align}
Unlike for the RR case, here we have to indicate the causal `$i0$' prescriptions explicitly for the propagators depending on the virtual momentum $k_2^\mu$, because they determine the sign of the imaginary part of the one-loop virtual correction. To build the DE system in $\epsilon$-form, we convert the set of MIs to the UT basis as follow
\begin{align}
\vec{\cI}&^{\rm RV}=\Big\{
\frac{\epsilon y }{x z}I_{2,0,1,1,0,0,0,2}^{{\rm RV}1} \,,
\frac{\epsilon^2 \sqrt{z^2-1}}{z}I_{1,0,1,2,0,0,0,1}^{{\rm RV}1} \,,
\frac{\epsilon z}{x y} I_{2,0,1,1,0,2,0,0}^{{\rm RV}1} \,,
\frac{\epsilon^2 \sqrt{y^2-1}}{y}  I_{1,0,1,2,0,1,0,0}^{{\rm RV}1} \,,
\nn\\&
\frac{\epsilon^2 \sqrt{z^2-1}}{x y} I_{1,0,1,1,0,2,1,0}^{{\rm RV}1} \,,
\frac{\epsilon^2 \sqrt{x^2-1}}{x} I_{1,0,1,2,0,1,1,0}^{{\rm RV}1} \,,
\frac{\epsilon^2 \sqrt{y^2-1}}{x z} I_{1,0,1,1,1,0,0,2}^{{\rm RV}1} \,,
\frac{\epsilon^2 \sqrt{x^2-1}}{x}  I_{1,0,1,2,1,0,0,1}^{{\rm RV}1} \,,
\nn\\&
\frac{\epsilon^2 \sqrt{x^2-1}}{x} I_{2,0,1,1,0,1,0,1}^{{\rm RV}2} \,,
\frac{\epsilon^2 \sqrt{x^2-1} \sqrt{y^2-1}}{x y} I_{1,0,1,1,0,2,0,1}^{{\rm RV}2} \,,
\frac{\epsilon^2 \sqrt{x^2-1}}{x} I_{2,1,1,0,0,1,0,1}^{{\rm RV}2} \,,
\nn\\&
\frac{\epsilon^2 \sqrt{x^2-1} \sqrt{z^2-1}}{x z} I_{1,1,1,0,0,1,0,2}^{{\rm RV}2} \,,
\frac{\epsilon^3 \sqrt{x^2-1} \sqrt{y^2-1}}{x y} I_{1,1,1,0,1,1,0,1}^{{\rm RV}2} \,,
\epsilon^2 \left(1-1/x^2\right) I_{1,1,1,0,1,1,0,2}^{{\rm RV}2} \,,
\nn\\&
\frac{\epsilon^3 \sqrt{x^2-1}}{x} I_{1,1,1,1,0,1,0,1}^{{\rm RV}2} \,,
\frac{\epsilon^3 \sqrt{y^2-1}}{y} I_{1,1,1,1,0,1,-1,1}^{{\rm RV}2}
-\frac{\epsilon^3 \sqrt{y^2-1}}{y} I_{1,1,1,1,0,1,0,0}^{{\rm RV}2} \,,
\nn\\&
\frac{\epsilon^3 \sqrt{z^2-1}}{z} I_{1,1,1,1,-1,1,0,1}^{{\rm RV}2}
-\frac{\epsilon^3 \sqrt{z^2-1}}{z} I_{1,1,1,1,0,0,0,1}^{{\rm RV}2} \,,
8 \epsilon^3 I_{1,1,1,1,-1,1,0,1}^{{\rm RV}2}-8 \epsilon^3 I_{1,1,1,1,0,0,0,1}^{{\rm RV}2}
\nn\\& \qquad
+2 \epsilon^3 I_{1,1,1,1,0,1,0,1}^{{\rm RV}2}+\frac{\epsilon^2 y}{x z} I_{1,1,1,1,-1,1,0,2}^{{\rm RV}2}
-\frac{\epsilon^2 y}{x z}  I_{1,1,1,1,0,0,0,2}^{{\rm RV}2} - 4 \epsilon^2 I_{1,0,1,2,0,0,0,1}^{{\rm RV}1} \,,
\nn\\& 
\frac{\epsilon^3 \sqrt{x^2-1} \sqrt{z^2-1}}{x z}  I_{1,1,1,0,0,1,1,1}^{{\rm RV}3} \,,
\epsilon^2 \left(1-1/x^2\right)  I_{1,1,1,0,0,2,1,1}^{{\rm RV}3}
\Big\}\,.
\end{align}
The boundary conditions are also determined at $v_1^\mu=v_2^\mu=\eta^\mu$. However, Coulomb singularities are encountered at this boundary point, because of the interaction between the two massive quarks at rest via the exchange of virtual gluons. Therefore, we have to resolve the singular behavior of the MIs in the limit of $x\to 1$ (i.e. $r\to 0$) before expanding them in $\epsilon$. As a result, we obtain the following analytical expressions for the boundary conditions
\begin{align}
\cI^{\rm RV}_{1,3}\Big|_{x=y=z=1} & = -8-16 i \pi  \epsilon +\frac{44 \pi ^2 \epsilon ^2}{3}+\left(\frac{112 \zeta_3}{3}+8 i \pi^3\right) \epsilon ^3+\left(-\frac{127 \pi ^4}{45}+\frac{224 i \pi }{3} \zeta_3\right) \epsilon ^4 +\cO(\epsilon^5) \,, \nn\\
\cI^{\rm RV}_{9,11}\Big|_{x=y=z=1} & = -16\, i \pi \, (2r)^{2 \epsilon }\,   e^{i \pi \epsilon } e^{2 \epsilon \gamma_E } \, \epsilon ^2\, \Gamma (2 \epsilon )\,\\
\cI^{\rm RV}_{i}\Big|_{x=y=z=1} & = 0\qquad \mbox{for}~ i\neq 1,3,9 ~\mbox{and}~ 11 \,.\nn
\end{align}
These boundary conditions have been cross checked with the results in ref.~\cite{Bierenbaum:2011gg}. Finally, by solving the DE equations order by order in $\epsilon$, the MIs can be expressed by GPLs with the same letters as for the NLO MIs.
As a further test of the NNLO calculation, we calculate all RR and RV MIs using the package \texttt{AMFlow}~\cite{Liu:2022chg} with high numerical precision at random benchmark points, finding complete agreement with the analytic calculation.

%%%%%%%%%%%%%%%%%%%%%%%%%%%%%%%%%%%%%%%%%%%%%%%
\section{Renormalisation and final result}\label{sec:renorm}
With the calculation outlined in the previous sections, the two-loop bare soft function can be expressed as
\begin{align}\label{eq:bsf2l}
S(\omega;\epsilon) = &\delta(\omega) 
+\frac{Z_\alpha \alpha_s}{4\pi}\frac{1}{\omega}\left(\frac{\mu}{\hat \omega}\right)^{2\epsilon} C_F K_F(x,y,z) \nn\\
&+\left(\frac{Z_\alpha \alpha_s}{4\pi}\right)^2 \frac{1}{\omega}\left(\frac{\mu}{\hat \omega}\right)^{4\epsilon} 
C_F\left[C_F K_{FF}(x,y,z) + C_A K_{FA}(x,y,z)+ n_f T_F K_{Ff}(x,y,z)\right]\,,
\end{align}
where $\hat{\omega} = \omega/\sqrt{\eta^2/4}$, and $Z_\alpha \alpha_s \mu^{2\epsilon}$ gives the bare QCD coupling constant with
\begin{equation}
Z_\alpha=1+\frac{\alpha_s}{4\pi}\left(-\frac{\beta_0}{\epsilon}\right)
+\left(\frac{\alpha_s}{4\pi}\right)^2\left(\frac{\beta_0^2}{\epsilon^2}
-\frac{\beta_1}{2\epsilon}\right)
+{\cal O}\left(\alpha_s^3\right)\,.
\label{eq:Zalpha}
\end{equation}
The NLO coefficient $K_F(x,y,z)$ is given by
\begin{align}
K_F=&-\frac{8 x}{\sqrt{x^2-1}}G_{1,r}-8
+8 \epsilon \Bigg( \frac{y }{\sqrt{y^2-1}}G_{1,t}+\frac{z }{\sqrt{z^2-1}}G_{1,u}
\nn\\
&\qquad -\frac{x}{\sqrt{x^2-1}}\Big[G_{\lambda _1,1,r}+G_{\lambda _2,1,r}+G_{\lambda _3,1,r}+G_{\lambda _4,1,r}-2 G_{0,1,r}-2 G_{2,1,r}
\nn\\
&\qquad\quad  +G_{1,t} \left(G_{\lambda _1,r}-G_{\lambda _2,r}-G_{\lambda _3,r}+G_{\lambda _4,r}\right)+G_{1,u} \left(G_{\lambda _1,r}-G_{\lambda _2,r}+G_{\lambda _3,r}-G_{\lambda _4,r}\right)\Big]
\Bigg)
\nn\\
&+2  \epsilon^2 \Bigg(
\pi ^2+\frac{8 y}{\sqrt{y^2-1}}\left(G_{0,1,t}-G_{1,1,t}+G_{2,1,t}\right) + \frac{8 z}{\sqrt{z^2-1}}\left(G_{0,1,u}-G_{1,1,u}+G_{2,1,u}\right)
\nn\\
&\quad
+\frac{x}{\sqrt{x^2-1}}\Big[\pi ^2 G_{1,r}-8 \left(G_{\lambda _1,r}-G_{\lambda _2,r}-G_{\lambda _3,r}+G_{\lambda _4,r}\right)\left(G_{0,1,t}-G_{1,1,t}+G_{2,1,t}\right)
\nn\\
&\qquad
-8 \left(G_{\lambda _1,r}-G_{\lambda _2,r}+G_{\lambda _3,r}-G_{\lambda _4,r}\right) \left(G_{0,1,u}-G_{1,1,u}+G_{2,1,u}\right)
\nn\\
&\qquad
+4 G_{1,t} \Big(2 G_{0,\lambda_1,r}-2 G_{0,\lambda _2,r}-2 G_{0,\lambda _3,r}+2 G_{0,\lambda _4,r}+2 G_{2,\lambda _1,r}-2 G_{2,\lambda _2,r}
\nn\\
&\qquad\quad 
-2 G_{2,\lambda _3,r}+2 G_{2,\lambda _4,r}-G_{\lambda _1,\lambda _1,r}+G_{\lambda _1,\lambda_2,r}+G_{\lambda _1,\lambda _3,r}-G_{\lambda _1,\lambda _4,r}
\nn\\
&\qquad\quad
-G_{\lambda _2,\lambda _1,r}+G_{\lambda _2,\lambda _2,r}+G_{\lambda _2,\lambda _3,r}-G_{\lambda _2,\lambda _4,r}-G_{\lambda _3,\lambda_1,r}+G_{\lambda _3,\lambda _2,r}
\nn\\
&\qquad\quad
+G_{\lambda _3,\lambda _3,r}-G_{\lambda _3,\lambda _4,r}-G_{\lambda _4,\lambda _1,r}+G_{\lambda _4,\lambda _2,r}+G_{\lambda _4,\lambda _3,r}-G_{\lambda _4,\lambda_4,r}\Big)
\nn\\
&\qquad
+4 G_{1,u} \Big(2 G_{0,\lambda _1,r}-2 G_{0,\lambda _2,r}+2 G_{0,\lambda _3,r}-2 G_{0,\lambda _4,r}+2 G_{2,\lambda _1,r}-2 G_{2,\lambda _2,r}
\nn\\
&\qquad\quad
+2 G_{2,\lambda _3,r}-2 G_{2,\lambda_4,r}-G_{\lambda _1,\lambda _1,r}+G_{\lambda _1,\lambda _2,r}-G_{\lambda _1,\lambda _3,r}+G_{\lambda _1,\lambda _4,r}
\nn\\
&\qquad\quad
-G_{\lambda _2,\lambda _1,r}+G_{\lambda _2,\lambda _2,r}-G_{\lambda _2,\lambda_3,r}+G_{\lambda _2,\lambda _4,r}-G_{\lambda _3,\lambda _1,r}+G_{\lambda _3,\lambda _2,r}
\nn\\
&\qquad\quad
-G_{\lambda _3,\lambda _3,r}+G_{\lambda _3,\lambda _4,r}-G_{\lambda _4,\lambda _1,r}+G_{\lambda _4,\lambda _2,r}-G_{\lambda _4,\lambda _3,r}+G_{\lambda _4,\lambda _4,r}\Big)
\nn\\
&\qquad
-4 \Big(4 G_{0,0,1,r}+4 G_{0,2,1,r}-2 G_{0,\lambda _1,1,r}-2 G_{0,\lambda _2,1,r}-2 G_{0,\lambda _3,1,r}-2 G_{0,\lambda _4,1,r}
\nn\\
&\qquad\quad
+4 G_{2,0,1,r}+4 G_{2,2,1,r}-2 G_{2,\lambda _1,1,r}-2 G_{2,\lambda _2,1,r}-2 G_{2,\lambda _3,1,r}-2 G_{2,\lambda _4,1,r}
\nn\\
&\qquad\quad
-2 G_{\lambda _1,0,1,r}-2 G_{\lambda _1,2,1,r}+G_{\lambda _1,\lambda_1,1,r}+G_{\lambda _1,\lambda _2,1,r}+G_{\lambda _1,\lambda _3,1,r}+G_{\lambda _1,\lambda _4,1,r}
\nn\\
&\qquad\quad
-2 G_{\lambda _2,0,1,r}-2 G_{\lambda _2,2,1,r}+G_{\lambda _2,\lambda _1,1,r}+G_{\lambda _2,\lambda_2,1,r}+G_{\lambda _2,\lambda _3,1,r}+G_{\lambda _2,\lambda _4,1,r}
\nn\\
&\qquad\quad
-2 G_{\lambda _3,0,1,r}-2 G_{\lambda _3,2,1,r}+G_{\lambda _3,\lambda _1,1,r}+G_{\lambda _3,\lambda _2,1,r}+G_{\lambda _3,\lambda _3,1,r}+G_{\lambda _3,\lambda _4,1,r}
\nn\\
&\qquad\quad
-2 G_{\lambda _4,0,1,r}-2 G_{\lambda _4,2,1,r}+G_{\lambda _4,\lambda _1,1,r}+G_{\lambda _4,\lambda _2,1,r}+G_{\lambda _4,\lambda _3,1,r}+G_{\lambda _4,\lambda _4,1,r}\Big)\Big]
\Bigg)
\nn\\
&
+O\left(\epsilon ^3\right)\,,
\end{align}
where we denoted $G_{a_1,\dots,a_n,x} = G(a_1,\dots,a_n;x)$ for brevity.
The UV renormalisation of the soft function is more conveniently performed in Laplace space, because the Laplace transform
\begin{align}
\tilde{S}(L;\epsilon) \equiv\int_0^\infty d\omega\, \exp \left(-\frac{\omega}{s\, e^{\gamma_E}}\right) S(\omega;\epsilon)\,,\qquad 
\mbox{with} \quad L=\ln\left(\frac{2s}{\mu\sqrt{\eta^2}}\right)\,,
\end{align}
turns the momentum-space convolutions between renormalisation factor and soft function into local products. 
One can show that, thanks to the correspondence of $S(\omega;\epsilon)$ with the threshold and TMD soft functions discussed in Sec.~\ref{sec:smom}, the renormalisation of $\tilde{S}(L)$ is multiplicative (local). A simple way of seeing this is to boost the fully differential soft function into the rest frame of the $\eta^\mu$ vector, hence highlighting the direct correspondence with the threshold soft function, whose renormalisation is multiplicative in Laplace space.

The Laplace transform of the bare soft function in Eq.~(\ref{eq:bsf2l}) can be achieved straightforwardly by the replacement
\begin{equation}
\frac{1}{\omega}\left(\frac{\mu}{\hat \omega}\right)^{n\epsilon}\to 
e^{-n\epsilon(L+\gamma_E)}\Gamma(-n\epsilon)\,.
\end{equation}
The ultraviolet (UV) poles in the bare soft function can be absorbed into the renormalisation factor defined in the $\overline{\rm MS}$ scheme
\begin{align}\label{eq:renormaliationdef}
\tilde{S}(L;\epsilon) = Z_{S}(\epsilon,\mu) \,\tilde{S}(L,\mu)\,.
\end{align}
The RG equation of the soft function is given by
\begin{align}\label{eq:sRGE}
\frac{d}{d\ln\mu}\tilde{S}(L,\mu) = \Gamma^S (\alpha_s)\,\tilde{S}(L,\mu)\,.
\end{align}
The RG equation of the hard function in Eq.~\eqref{eq:xfac} takes the same form for RG consistency.

\subsection{Anomalous dimension of the fully differential soft function}\label{subsec:anomdim}
Since the cross section is free of singularities, the UV poles of the bare soft function have to cancel against the infrared (IR) poles of the hard function ${\cal H}(\{q_1,q_2,q_V\};\epsilon)$, which are governed by the anomalous dimension~\cite{Korchemsky:1987wg,Korchemsky:1991zp,Kidonakis:2009ev,Becher:2009kw}
\begin{align}\label{eq:gaQQ}
\Gamma_{QQ}(\{v_1,v_2\},\alpha_s)=C_F\gamma_{\rm cusp}(\beta,\alpha_s) + 2 \gamma^Q(\alpha_s)\,,
\end{align}
where $\gamma_{\rm cusp}(\beta,\alpha_s)$ is the massive cusp anomalous dimension 
\begin{align}
\gamma_{\rm{ cusp}}(\beta,\alpha_s) = &
\gamma_{\rm{cusp}}(\alpha_s) \beta\coth \beta \nn \\
&+ \frac{C_A}{2}\left(\frac{\alpha_s}{\pi}\right)^2 
\Bigg[\left(\text{Li}_3\left(e^{-2 \beta }\right)+\beta\text{Li}_2\left(e^{-2 \beta }\right)+\frac{\beta ^3}{3}+\frac{\pi^2}{6}\beta -\zeta_3\right) \coth^2\beta  \nn\\
&\qquad +\left(\text{Li}_2\left(e^{-2 \beta}\right) -2 \beta\ln\left(1-e^{-2 \beta }\right) 
-\frac{\pi^2}{6} (\beta +1) -\beta^2 - \frac{\beta ^3}{3}\right) \coth\beta  \nn\\
&\qquad+\zeta_3 + \frac{\pi^2}{6} + \beta ^2\Bigg] + {\cal O}(\alpha_s^3) \,.
\end{align}
with the cusp angle defined by $\beta = \cosh^{-1}(-x) = -\ln(x - \sqrt{x^2-1})-i\pi$\,, and $\gamma^Q(\alpha_s)$ is the soft anomalous dimension for heavy quarks. The coefficients of $\gamma^Q(\alpha_s)$ and the massless cusp anomalous dimension $\gamma_{\rm cusp}(\alpha_s)$ are presented in Appendix~\ref{app:anomdims} up to two-loop order. RG invariance of the cross section then implies that
\begin{align}
\Gamma^S (\alpha_s) = -\Gamma^H (\alpha_s) = -2\, {\rm Re}\, \Gamma_{QQ}(\{v_1,v_2\},\alpha_s)\,,
\end{align}
which determines the renormalisation factor
\begin{align}
Z_S(\epsilon,\mu) ={}& 1 + \frac{\alpha_s}{4\pi}\frac{\Gamma^S_0}{2\epsilon} 
%\nn\\&
+ \left( \frac{\alpha_s}{4\pi} \right)^2 \! \left[
 \frac{\Gamma^S_0}{8\epsilon^2} 
\left( \Gamma^S_0 -2\beta_0 \right) 
+ \frac{\Gamma^S_1}{4\epsilon} \right] + {\cal O}(\alpha_s^3)\,,
\label{eq:ZsNNLO}
\end{align}
where we used the conventions
\begin{align}
 \Gamma^S(\alpha_s)&=\sum_{n=0}^{\infty} \Gamma_n^S
 \left(\frac{\alpha_s}{4\pi}\right)^{\!n+1}
\qquad\mbox{and}
\qquad \beta(\alpha_s) =-2\alpha_s\sum_{n=0}^{\infty}\beta_n
\left(\frac{\alpha_s}{4\pi}\right)^{\!n+1}\,.
\label{eq:anomdimexp}
\end{align}

%%%%%%%%%%%%%%%%%%%%%%%%%%%%%%%%%%%%%%%%%%
\subsection{Renormalised results}\label{subsec:renres}
%%%%%%%%%%%%%%%%%%%%%%%%%%%%%%%%%%%%%%%%%%
As expected, $Z_s$ in Eq.~(\ref{eq:ZsNNLO}) absorbs all the divergences of the bare soft function $\tilde{S}(L;\epsilon)$ up to two-loop order. This represents a strong check of our calculation. As a solution to the RG equation in Eq.~(\ref{eq:sRGE}), the renormalised soft function in Laplace space can be expressed as 
\begin{align}
\tilde{S}(L,\mu)={}&1+\frac{\alpha_s}{4\pi}\left(-\Gamma^S_0 L +c^S_1\right) \nn\\
&\quad
+\left(\frac{\alpha_s}{4\pi}\right)^2\Bigg[ \left(\beta _0 \Gamma_0^S
+\frac{1}{2}\left(\Gamma_0^{S}\right)^2\right) L^2
-\left[c_1^S \left(\Gamma _0^S+2 \beta _0\right)+\Gamma_1^S\right]L + c^S_2\Bigg]
\,,
\label{eq:finalrensoft}
\end{align}
where the one-loop constant coefficient is given by
\begin{align}
c_1^s = &4 C_F \Bigg( -\frac{y }{\sqrt{y^2-1}}G_{1,t}  - \frac{z }{\sqrt{z^2-1}}G_{1,u}
\nn\\
&\qquad +\frac{x}{\sqrt{x^2-1}}\Big[G_{\lambda _1,1,r}+G_{\lambda _2,1,r}+G_{\lambda _3,1,r}+G_{\lambda _4,1,r}-2 G_{0,1,r}-2 G_{2,1,r}
\nn\\
&\qquad\quad  +G_{1,t} \left(G_{\lambda _1,r}-G_{\lambda _2,r}-G_{\lambda _3,r}+G_{\lambda _4,r}\right)+G_{1,u} \left(G_{\lambda _1,r}-G_{\lambda _2,r}+G_{\lambda _3,r}-G_{\lambda _4,r}\right)\Big]
\Bigg)\,.
\end{align}
The expression of $c_2^S$ is too long to be reported here, so it is provided as an ancillary file with this article.

As a final check of our results, we consider the two-loop threshold soft function which can be obtained from the fully differential soft function following Eqs.~\eqref{eq:threshold-soft},~\eqref{eq:sxtop},~\eqref{eq:threshold-case}. In particular, we specialise to the case of $Q\bar{Q}$ production (i.e. without a colour singlet $V$), for which the two-loop threshold soft function was previously calculated in Ref.~\cite{vonManteuffel:2014mva}. Our test reproduces the results of the latter reference, providing a strong check of our calculation.

To obtain the TMD soft function in position space described by Eqs.~(\ref{eq:sxtop}) and (\ref{eq:tmd-case}), we can see that it is equivalent to the soft function in Laplace space just by replacing $s\to e^{-\gamma_E}/x_T$, i.e.  
\begin{align}
L\to L_\perp = \ln\left(\frac{2 e^{-\gamma_E}}{x_T\, \mu}\right)\,.
\end{align}

\section{Conclusions and outlook}\label{sec:conclu}
In this article, we presented the first calculation of the fully differential soft function for the production of a pair of heavy quarks accompanied by a generic colour singlet system. The calculation is performed in generic final state kinematics and retaining differential information in the total momentum of the real QCD radiation.
The resulting soft function can be mapped onto soft functions entering a number of factorisation theorems at N$^3$LL order, for instance that for the cross section for $Q\bar{Q}V$ production at threshold and the TMD soft function which is relevant for observables which depend on the total transverse momentum of the final-state radiation.
As such, our calculation is also a central ingredient for the two-loop fully differential soft function for $Q\bar{Q}V$ production at hadron colliders, of which it constitutes the contribution from final-final dipoles.
Besides the relevance in the context of resummed calculations, the results of this article are instrumental for the implementation of slicing-based subtraction schemes for NNLO calculations of collider observables, as well as their matching to parton showers.

We provide analytic results for the soft function, obtained by mapping the problem into one that can be handled using modern Feynman integrals techniques directly in momentum space. The final result is provided in electronic form as an ancillary file to this article, expressed in terms of Goncharov polylogarithms and can be used for an arbitrary-precision numerical evaluation.
Our work also opens to the possibility to calculate the fully differential soft function for $Q\bar{Q}V$ at hadron colliders, which could be approached with the same method presented in this article. We leave this, together with phenomenological applications of the results presented here, to future work.

\begin{acknowledgments}
We are grateful to Xiao Liu for his help with the use of the \texttt{AMFlow} package for the evaluation of integrals with cut linear propagators, and to Luca Buonocore and Daniele Lombardi for several constructive discussions and collaboration on projects related to the topics of this article. 
We thank the Munich Institute for Astro-, Particle and BioPhysics (MIAPbP) for hospitality during the final stages of this work.
Z.L.L is supported by Institute of High Energy Physics (IHEP, CAS) under Grant No. E35159U1. P.F.M is funded by the European Union (ERC,
grant agreement No. 101044599). Views and opinions
expressed are however those of the authors only and do
not necessarily reflect those of the European Union or the
European Research Council Executive Agency. Neither
the European Union nor the granting authority can be
held responsible for them.
All graphs were drawn using \texttt{JaxoDraw}~\cite{Binosi:2008ig}.
\end{acknowledgments}

\appendix
\section{Two-loop anomalous dimensions}
\label{app:anomdims}
We adopt the following notation for the expansion of anomalous dimensions in Eq.~(\ref{eq:gaQQ})
\begin{align}%\label{eq:anomdimexp}
 \gamma_{\rm cusp}(\alpha_s)&=\sum_{n=0}^{\infty} \gamma_n^{\rm cusp}
 \left(\frac{\alpha_s}{4\pi}\right)^{\!n+1}\,,
\qquad \gamma^Q(\alpha_s) =\sum_{n=0}^{\infty}\gamma^Q_n
\left(\frac{\alpha_s}{4\pi}\right)^{\!n+1}\,.
\end{align}
The coefficients of the massless cusp anomalous dimension up to two-loop order are~\cite{Korchemsky:1987wg}
\begin{align}
\gamma_0^{\rm cusp} ={}& 4 \,, \nn\\
\gamma_1^{\rm cusp} ={}& \biggr( \frac{268}{9} 
- \frac{4\pi^2}{3} \biggl) C_A - \frac{80}{9}\,T_F n_f \,.
\end{align}
The coefficients up to four-loop order have been obtained in refs.~\cite{Moch:2004pa,Vogt:2004mw,Henn:2019swt,vonManteuffel:2020vjv}.
The coefficients of the anomalous dimension for external massive quark up to two-loop order are
\begin{align}
\gamma_0^Q ={}& -2C_F \,, \label{eq:gammaQ0} \\ 
\gamma_1^Q={}& 
C_F \bigg[C_A\biggl(\frac{2\pi^2}{3}-\frac{98}{9}-4\zeta_3\biggr)
+n_f T_F \frac{40}{9}  \bigg] \,. \label{eq:gammaQ1}
\end{align}
The coefficient at three-loop order is available in refs.~\cite{Grozin:2014hna,Bruser:2019yjk}.

\phantomsection
\addcontentsline{toc}{section}{References}
\bibliographystyle{jhep}
\bibliography{refs}

\end{document}